\documentclass[submission, Phys]{SciPost}

\usepackage{revsymb}
\hypersetup{
    colorlinks=red,
    citecolor=blue,
    filecolor=magenta,      
    urlcolor=red,
}

\begin{document}

\begin{center}{\Large \textbf{
$cos(2\phi_h)$ asymmetry in $J/\psi$ production  in unpolarized $ep$ collision in NRQCD
}}\end{center}

\begin{center}
Raj Kishore\textsuperscript{1},
Asmita Mukherjee\textsuperscript{1},
Mariyah Siddiqah\textsuperscript{1*}
\end{center}

\begin{center}
{\bf 1} Indian Institute of Technology Bombay.
\\
* shah.siddiqah@gmail.com
\end{center}

\begin{center}
\end{center}

\vspace{-6mm}
\section*{Abstract}
{\bf
We present a recent calculation of the transverse momentum dependent gluon distributions inside unpolarised protons and show how the ratio of the linearly polarized and the unpolarized gluon
distribution in the proton can be probed by looking at $Cos(2\phi_h)$ asymmetry in $J/\psi$ production  in unpolarized $ep$ collision. We use NRQCD for estimating  $J/\psi$ production and include contribution both from color singlet and color octet states.
}


\vspace{-6mm}
\section{Introduction}
\label{sec:intro}
Transverse momentum dependent (TMD) parton distribution functions (PDFs) play an important role in understanding the spin
and spatial structure of the proton. They not only provide the information about the longitudinal momentum fraction but also give us information about the internal transverse momentum $k_{\perp}$ carried by partons. Unlike collinear PDFs, which are universal, TMDs are
 process dependent due to their initial and final state interactions.
The two experiments where TMDs can be extracted are SIDIS and DY.  In these experiments the observables of most interest are the single-spin
 asymmetries (SSA) and azimuthal asymmetries. 
The leading-twist distributions, unpolarised gluon TMD $ f_1^g(x,\mathbf{k}_{\perp}^2)$ and $h_1^{\perp g}(x,\mathbf{k}_{\perp}^2)$, the
so-called Boer–Mulders function, are greatly relevant
to these asymmetries. The function $f_1^g(x,\mathbf{k}_{\perp}^2)$ represents the probability of finding an unpolarized gluon, within an unpolarized hadron, with a longitudinal momentum fraction $x$ and transverse momentum $k_\perp$, while as $h_1^{\perp g}(x,\mathbf{k}_{\perp}^2)$ represents the distribution of linearly polarized gluons within the unpolarized hadron.\\
 Due mainly to the lack of experimental data, information on the  $h_1^{\perp g}(x,\mathbf{k}_{\perp}^2)$ is very limited. Heavy quark pair or dijet
production in SIDIS \cite{pisano2013linear}, diphoton pair \cite{qiu2011probing} and $\Upsilon (1S)$+jet \cite{den2014accessing} production in $pp$ collision have been suggested for extracting $h_1^{\perp g}(x,\mathbf{k}_{\perp}^2)$. It has been seen
in these processes that measuring azimuthal asymmetries, we can probe  $h_1^{\perp g}(x,\mathbf{k}_{\perp}^2)$.\\
  Quarkonium production is an important tool to probe the gluon TMDs  (see \cite{lansberg2020new}for a recent review).
  In Ref. \cite{mukherjee2017j} the authors probed $h_1^{\perp g}$ in $cos(2\phi_h)$ asymmetry in $J/\psi$ production through the leading order
(LO) process $\gamma^{\ast}+g\rightarrow J/\psi$ at the future EIC at $z=1$, where $z$ measures the fraction of photon energy transferred to $J/\psi$. This was extended to the region $z<1$ in \cite{kishore2019accessing}, where only the color singlet contribution within the non-relativistic QCD (NRQCD) was considered.  
In this article we review a recent calculation of the $cos (2 \phi_h)$ asymmetry in $J/\psi$ production in electron-proton collision\cite{kishore2021cos}.  The production of $J/\psi$ is calculated in the NRQCD framework with the inclusion of both color singlet and color octet contributions. We will be taking $\gamma^{\ast}+g\rightarrow J/\psi+g$ partonic subprocess into the consideration, and investigate mainly  the small-$x$ region, where the gluon TMDs play a major role. 

The rest of this paper is organized as follows. The analytical framework of our calculation is discussed in Section 2. In Section 3, we present our numerical results and finally, in Section 4, we conclude.
\vspace{-6mm}
\section{Azimuthal asymmetry in $J/\psi$ leptoproduction}
The $cos(2\phi_h)$ asymmetry for $e(l)+p(P)\rightarrow e(l')+J/\psi(P_h)+X(P_x)$ process is defined as 
\begin{eqnarray}
\langle cos(2\phi_h)\rangle&=&\frac{\int d\phi_h cos(2\phi_h)d\sigma}{\int d\phi_h d\sigma},
\end{eqnarray}
where $\phi_h$ is the azimuthal angle of $J/\psi$ production plane with the lepton plane and $d\sigma$ is the differential scattering cross section.
%
We consider the frame  in which the incoming proton and the virtual photon exchanged in the 
process move  in $+z$ and $-z$ directions.
The kinematics here is defined in terms of two light-like vectors with the help of a Sudakov decomposition, here chosen to
be the momentum $P(=n_{-})$ of the incoming proton, and a second vector $n(=n_{+})$, obeying the relations $n\cdot P = 1$ and $n_{+}^2=n_{-}^ 2 = 0$.
Since at small-$x$ the proton is rich in gluons, the dominant partonic sub-process at NLO for the $J/\psi$ production is $\gamma^{\ast}(q)+g(k)\rightarrow J/\psi(P_h)+g(p_g)$.
The differential scattering cross-section can be written as a convolution of leptonic tensor, a soft parton correlator for the incoming hadron and a hard part:
\begin{equation}
\begin{aligned}
 d\sigma={} &\frac{1}{2s}\frac{d^3l'}{(2\pi)^32E_l'}\frac{d^3P_{h}}{(2\pi)^32E_{P_h}}\int \frac{d^3p_g}{(2\pi)^32E_g}
  \int dxd^2\mathbf{k}_{\perp}(2\pi)^4\delta(q+k-P_h-p_g)\\
  &\times \frac{1}{Q^4}\mathcal{L}^{\mu\mu'}(l,q)\Phi^{\nu\nu'}
  (x,\mathbf{k}_{\perp})~\mathcal{M}_{\mu\nu}(\mathcal{M}_{\mu'\nu'})^{\ast}
  \label{ds}
 \end{aligned}
\end{equation}
  The term $\mathcal{M}_{\mu\nu}$ represents  the amplitude of $J/\psi$ production in  the ${\gamma^{\ast}+g\rightarrow J/\psi+g}$ partonic sub-process and $\mathcal{L}^{\mu\mu'}(l,q)$ is the leptonic tensor.
At leading twist, the gluon correlator of the unpolarized proton contains two TMD gluon distribution functions
\begin{eqnarray}
 \Phi_g^{\nu\nu'}(x,\mathbf{k}_{\perp})=-\frac{1}{2x}\bigg\{g_{\perp}^{\nu\nu'}f_1^g(x,\mathbf{k}_{\perp}^2)-\left(\frac{k_{\perp}^{\nu}k_{\perp}^{\nu'}}{M_p^2}+g_{\perp}^{\nu\nu'}\frac{\mathbf{k}_{\perp}^2}{2M_p^2}\right)h_1^{\perp g}(x,\mathbf{k}_{\perp}^2)\bigg\}.
\end{eqnarray}
Here $g_{\perp}^{\nu\nu'}=g^{\nu\nu'}-P^{\nu}n^{\nu'}/P\cdot n-P^{\nu'}n^{\nu}/P\cdot n$. 
Following Refs. \cite{kishore2021cos} and references within, to which we refer the reader for more details, we could write the final expression for the azimuthal asymmetry as: 
%
\begin{eqnarray}
 \langle cos(2\phi_h)\rangle\propto\frac{ \int k_\perp d k_\perp \Big (  A_2~f_1^g(x,\textbf{k}_{\perp}^2)+\frac{k_{\perp}^2}{M_p^2}~B_2~h_1^{\perp g}(x,\textbf{k}_{\perp}^2) \Big ) }{\int k_\perp d k_\perp \Big (A_0~f_1^g(x,\textbf{k}_{\perp}^2)+\frac{k_{\perp}^2}{M_p^2}~B_0~h_1^{\perp g}(x,\textbf{k}_{\perp}^2) \Big )}.
\end{eqnarray}
In the kinematics considered above we found that the unpolarized gluon distribution and the linearly polarized distributions are not disentangled, however the above result can be used for the extraction of their ratio. 

 \vspace{-7mm}

\section{ Results and discussion}

In this section, we present numerical estimates of the $cos(2\phi_h)$ asymmetry in the kinematical region to be accessed at EIC and to avoid the contribution from virtual diagrams.
We have also imposed a lower cutoff $z>0.1$ to avoid the gluon fragmentation contribution to $J/\psi$. We verified that changing the lower cutoff  does not affect the asymmetry much.
We took mass of the proton to be $M_p = 1$ GeV.   

The contraction in the calculation above for the different states  $i.e.,~{^1}S_0^{(8)}, ^3S_1^{(1,8)}$ and $^3P_{J(=0,1,2)}^{(8)}$ is calculated using the FeynCalc \cite{shtabovenko2020feyncalc,mertig1991feyn}. In all the plots of the asymmetry, the  
long distance matrix elements (LDMEs) are taken from Ref. \cite{chao2012j} except for the right panel of Fig.~\ref{g2}, where we have used different sets of LDMEs. 
We have used  two sets of parameterization for the TMDs to calculate the $cos(2\phi_h)$ asymmetry:~1) Gaussian-type parameterization
 \cite{boer2012polarized,mukherjee2017linearly}
and~ 2) McLerran-Venugopalan model
 \cite{mclerran1994computing,mclerran1994gluon,mclerran1994green} 
at small-$x$ region.  
\subsubsection{$cos(2\phi_h)$ asymmetry in the Gaussian parameterization}
The results for the Gaussian are obtained for $\sqrt{s}=100$ GeV and $Q^2=15$ GeV. We have incorporated only few of the results here, the reader could find the more details in Ref. \cite{kishore2021cos}.
\begin{figure}[h]
 
 \centering
 \includegraphics[width=.50\textwidth]{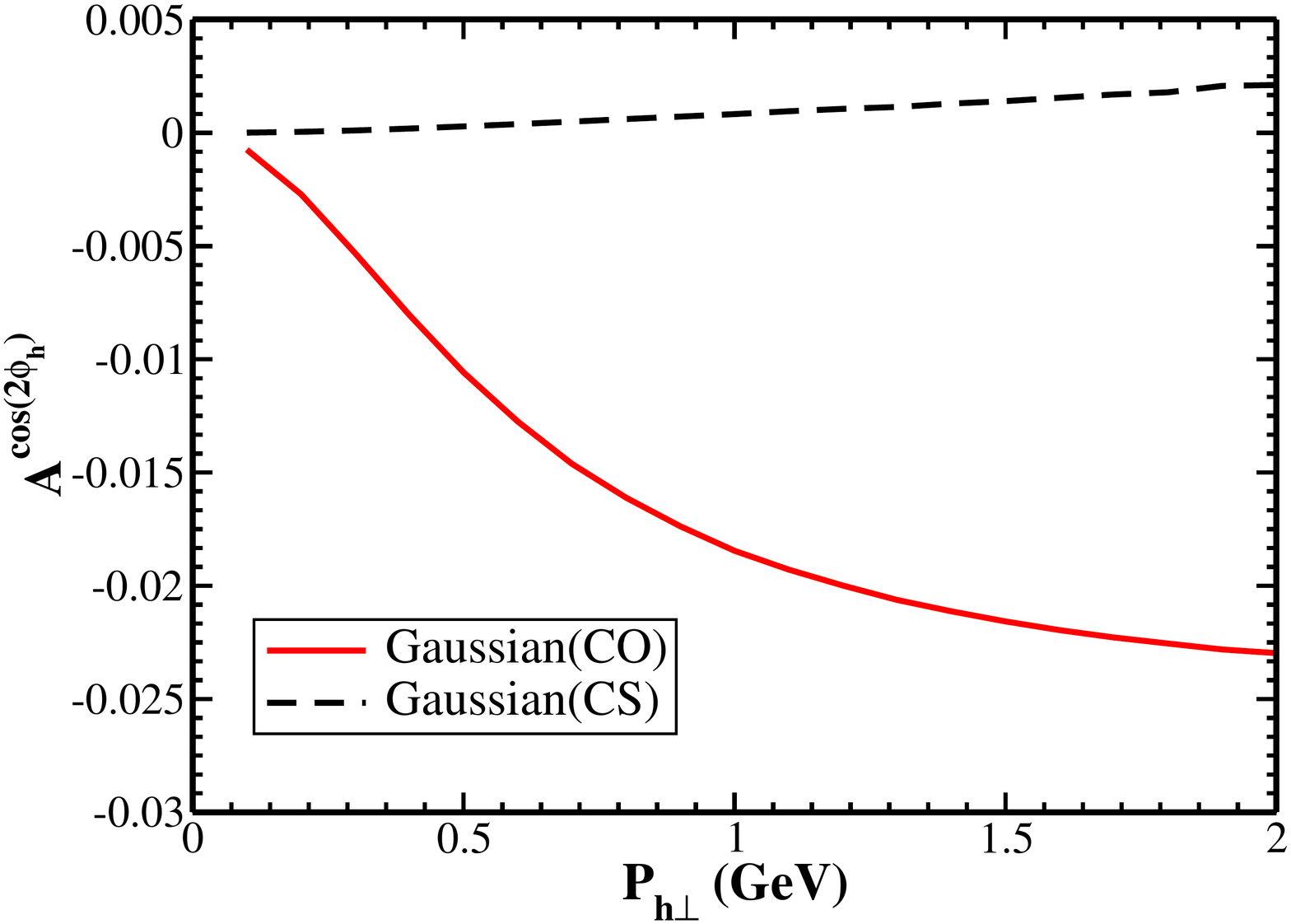}\hfill
 \includegraphics[width=.50\textwidth]{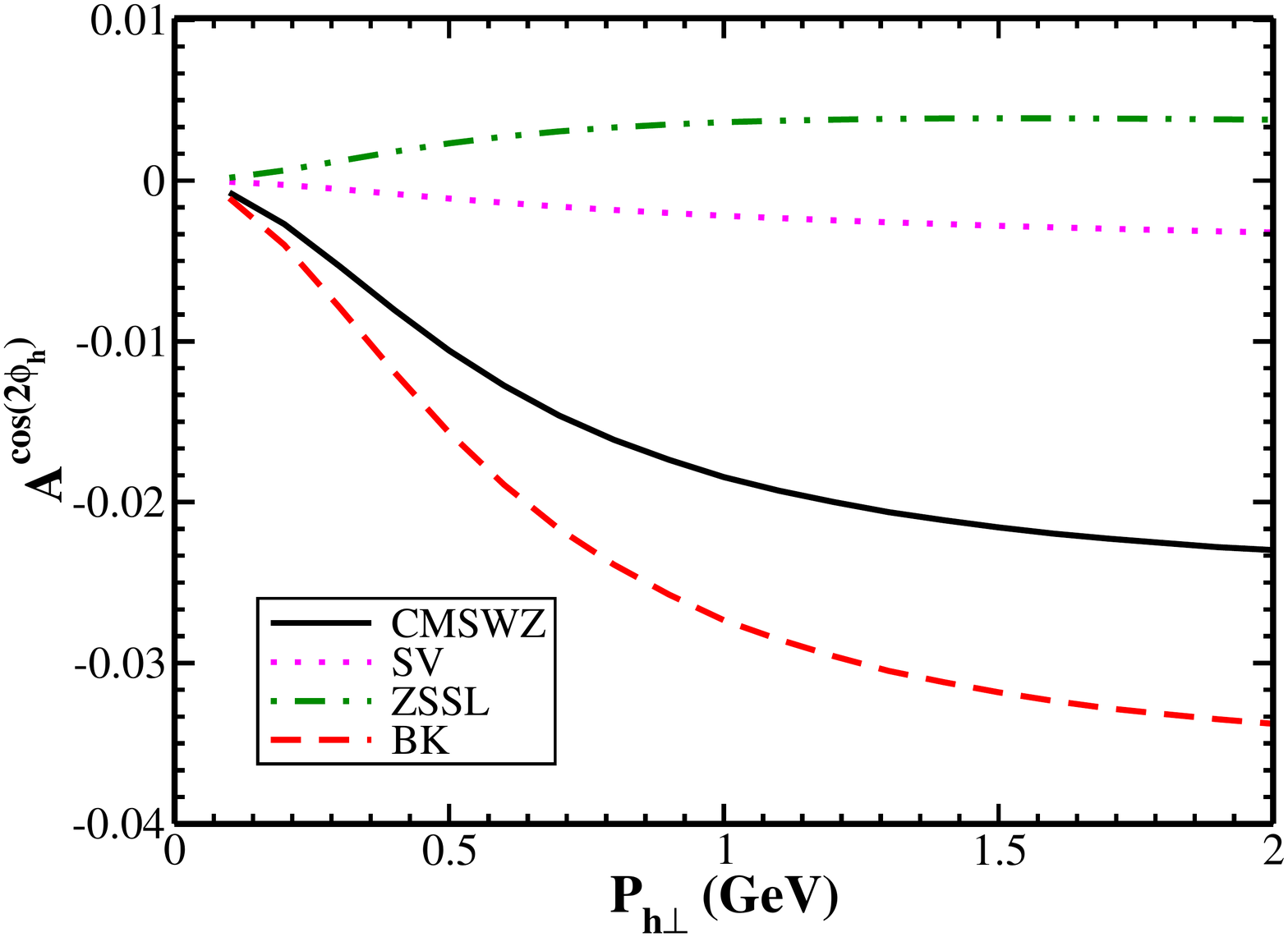}\hfill

 \caption{$cos(2\phi_h)$ asymmetry  in $e+p\rightarrow e+ J/\psi +X$
		process as function of $P_{h\perp}$ at $\sqrt{s}=100$ GeV and $Q^2=15$ GeV$^2$.
  		Left:  contributions from color singlet and color octet states  using the LDMEs set CMSWZ \cite{chao2012j}. Right: comparing the asymmetry for  different LDMEs sets: CMSWZ \cite{chao2012j}, SV \cite{sharma2013high}, ZSSL \cite{zhang2015impact}, BK \cite{butenschoen2011world}.}
 \label{g2}
 \end{figure}
 In the left panel of Fig. \ref{g2}, we present the contribution to the asymmetry from  the color singlet(CS) and the color octet(CO) states. From the plot, we see that  the color octet states are  giving a significant contribution to the asymmetry, whereas the contribution from the CS is almost zero and slightly positive in higher $P_{h \perp}$ region. In the right panel of Fig. \ref{g2}, we showed the asymmetry for different sets of LDMEs. We see that the magnitude and the sign of the asymmetry depends on the set of LDMEs used. In fact, this is because of different states contributing to the asymmetry  depending on the LDMEs. We have a larger asymmetry for LDMEs set BK.
\subsubsection{$cos(2\phi_h)$ asymmetry in the McLerran-Venugopalan (MV) model}
\begin{figure}[h]
 
 \centering
 \includegraphics[width=.50\textwidth]{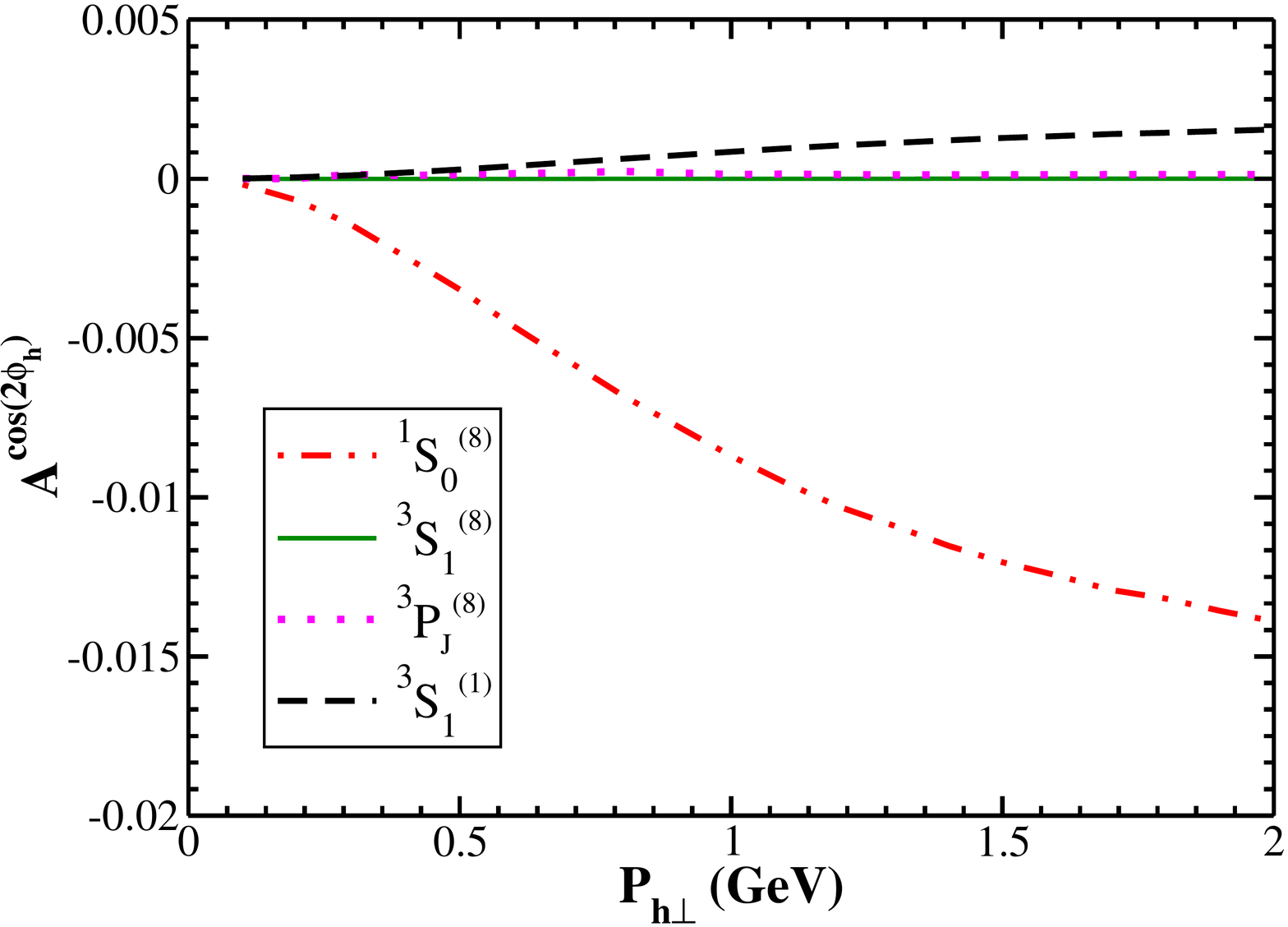}\hfill
   \includegraphics[width=.50\textwidth]{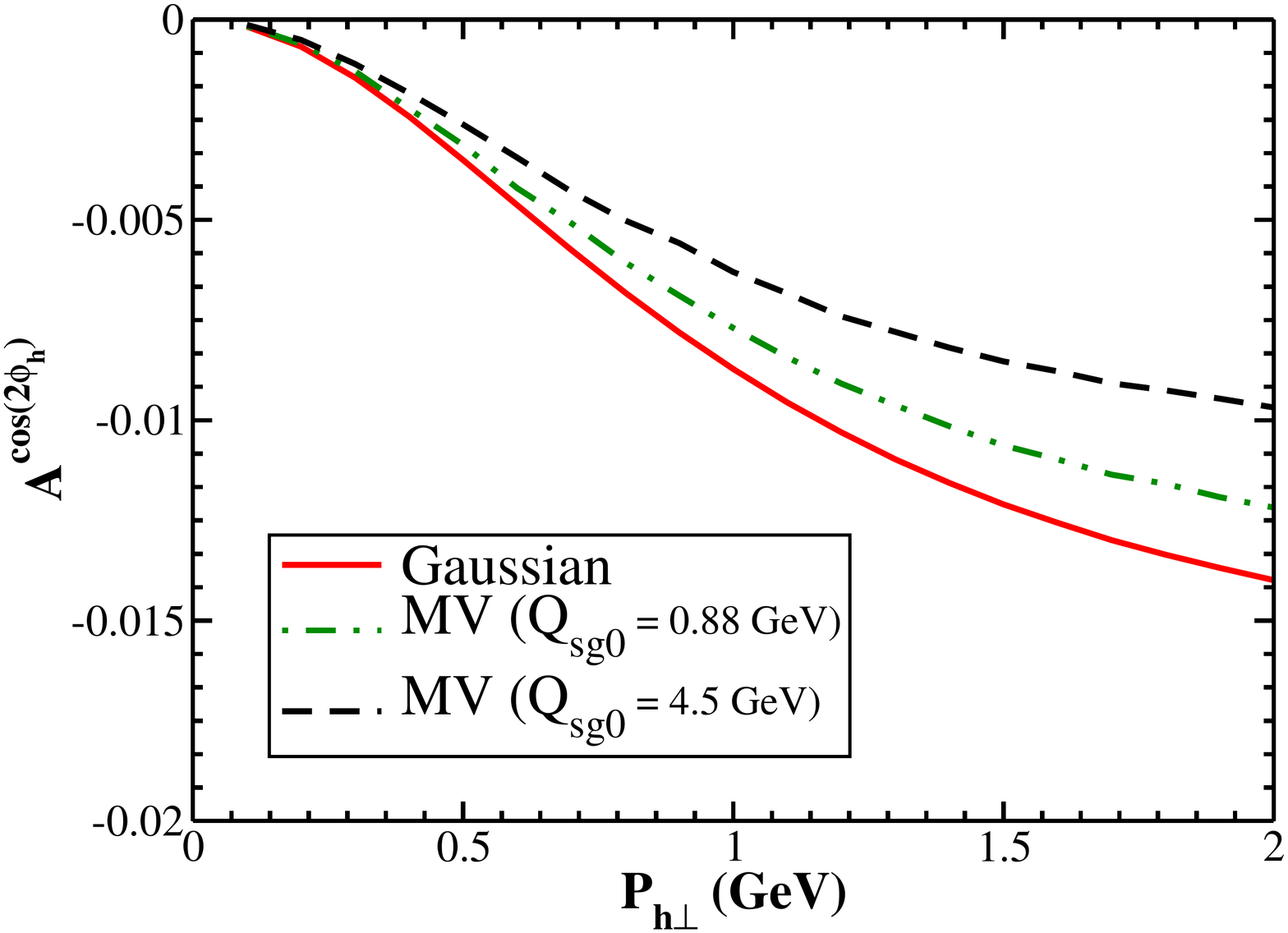}\hfill

 \caption{$cos(2\phi_h)$ asymmetry  in $e+p\rightarrow e+ J/\psi +X$
 		process as function of $P_{h\perp}$ at $\sqrt{s}=150$ GeV, $x=0.01$ and $z=0.7$. 
		Left: contribution to the $cos(2\phi)$ asymmetry coming from the individual states, as a function of $P_{h \perp}$ in the NRQCD framework using color octet model. 
        Right: comparison of the Gaussian and MV model (with two different values of $Q_{sg0}$). For both plots, CMSWZ set of LDMEs \cite{chao2012j} is used. }
 \label{mv2}

 \end{figure}

 For the MV parameterization,the results are obtained in the kinematic region defined by $\sqrt{s}=150$ GeV, $x=0.01$ and $z=0.7$. The integration ranges are $y\in[0.2,0.9]$ and $x_B\in[0.005,0.009]$. The value of $Q$ is set according to  $y,x_B$ and $s$.  In Fig. \ref{mv2} (Left), we present the contribution to the $cos(2\phi_h)$ asymmetry for the $J/\psi$ coming from the individual states, as a function of $P_{h \perp}$. The maximum contribution to the asymmetry comes from the  $^1S^{(8)}_0$ state. In the same Fig. \ref{mv2} (right), we show the  comparison of  the asymmetry in the Gaussian and MV model, respectively (with two different values of $Q_{sg0}$
 ) within the same kinematical region as discussed above.  The asymmetry in MV model depends on the saturation scale. We note that the Gaussian parameterization of the TMDs gives larger   $cos(2\phi_h)$ asymmetry.
 \vspace{-6mm}
\section{Conclusion}
We investigated  the $cos(2\phi_h)$  asymmetry in $J/\psi$ production in an electron-proton collision in the kinematics of the future EIC. We calculated them in the small-$x$ domain, where the gluon TMD, namely the unpolarized and linearly polarized gluon TMD are important in unpolarized scattering. The virtual-photon-gluon fusion process $\gamma^{\ast}+g\rightarrow J/\psi+g$ is the dominant sub-process in this kinematical region for $J/\psi$ production. We used the NRQCD-based color octet formalism to calculate the $J/\psi$ production rate.  A small but significant
$cos(2\phi_h)$ asymmetry is obtained.
We used the Gaussian and MV models for parameterization of TMDs.
The magnitude of the asymmetry was found to be larger with Gaussian parameterization. 
 We incorporated contributions both from CO and CS states.  
 The asymmetry depends on the LDMEs used; specifically, contributions from individual states were found to be significantly dependent on the set of LDMEs used. Overall, our results indicate  that the $cos(2 \phi_h)$ asymmetry in $J/\psi$ production might be a very useful tool for probing the ratio of the linearly polarized gluon TMD and the unpolarized gluon TMD  in the small-$x$ region at the EIC. 
  \vspace{-8mm}
\section{Acknowledgement}
M.Siddiqah thanks the organisers of DIS2021 for the opportunity to present this work.

\vspace{-6mm}
\nolinenumbers
 \bibliography{biblography_as.bib}

\end{document}